\def\beq{\begin{equation}}
\def\eeq{\end{equation}}
\def\bea{\begin{eqnarray}}
\def\eea{\end{eqnarray}}
\documentstyle[prl,preprint,aps]{revtex}
\begin{document}
\title{Inflationary solutions and inhomogeneous Kaluza-Klein cosmology
in ${\bf 4+n}$ dimensions}
\author{Santiago E. Perez Bergliaffa
\thanks{E-mail: bergliaf@venus.fisica.unlp.edu.ar}}
\address{Departamento de F\'{\i}sica, Universidad Nacional de La Plata\\
C.C. 67, 1900, La Plata, Buenos Aires, Argentina}
\maketitle
\thispagestyle{empty}

\begin{abstract}
We analyze the existence of inflationary solutions in an inhomogeneous 
Kaluza-Klein cosmological model in $4+n$ dimensions. It is 
shown 
that the 5-dimensional case is the exception rather than the rule,
in the sense that the system is 
integrable (under the assumption of the equation of state $p=k\rho$) for any
value of $k$. It is also shown that the cases $k=0$ and $k=1/3$ are integrable
if and only if $n=1$.
\end{abstract}
\vspace{1cm}
\noindent PACS {\em number(s):} 04.50.+h 98.80.Cq\\
\noindent Keywords: Multidimensional theories, inflationary solutions

\vspace{1cm}
Much of the recent work related to unification of the fundamental interactions
involve theories that are formulated in a spacetime with dimension $d>4$. 
Superstrings \cite{sustri} and M-theory \cite{mth} are examples of such 
theories. Although
with different motivations, the idea of a higher dimensional spacetime has been
advocated by different authors \cite{kle,dewit,crem,jac} since Kaluza 
\cite{kal} put forward a theory
that unified relativity and electromagnetism in a 5-dimensional spacetime. If
one admits that the extra dimensions have a physical reality (and are not 
a mere mathematical device to carry out the dimensional reduction of the theory,
as in \cite{crem}), then one must account for the fact that our universe is,
in the present era, manifestly 
4-dimensional. Usually, one adopts the view that the vacuum of the theory 
has undergone a spontaneous compactification \cite{cs2}, in which the extra 
dimensions
are compactified to the Planck size. However, it must be verified that
this assumption is consistent with cosmology. That is, there must exist 
cosmological solutions of the theory that, starting from a spacetime in 
which all dimensions had comparable size, exhibit the desirable feature of 
an unobservable internal space at later times. Solutions of this kind have been 
found for maximally symmetric spaces \cite{fre},  
homogeneous models \cite{abb,sah,chat1}, and also for
anisotropic models \cite{ber,ban}, but little attention has been paid to 
higher dimensional inhomogeneous models. These are important in the light of the 
findings of the COBE \cite{cobe}, 
which reveal the existence of inhomogeneities in
the early Universe. In turn, these might be accounted for through the 
existence of inhomogeneous extra dimensions in an early phase of the Universe
\cite{chat4}.

Several papers have been devoted to inhomogeneous Kaluza-Klein cosmology in the
case of one extra dimension \cite{chat4,chat2,chat3,ban2}. 
Here we deal with the more general case of 
an arbitrary number of extra dimensions in the presence of matter with a given 
equation of state.

We assume a flat and homogeneous 3-space, and we introduce the inhomogeneity
in the $n$ extra dimensions. The metric tensor is such that the 
$4+n$-dimensional interval takes the form
\beq
ds^2=dt^2 -R^2(t)(dr^2+r^2d\theta ^2+r^2\sin^2\theta d\phi^2) -A^2(r,t) dy^2 
\label{met}
\eeq
where $dy^2\equiv\sum_{i=4}^{4+n}dy_n^2$. The nonzero components of the 
stress-energy tensor are
\bea
T^0_{\;0} =\rho(r,t)+\Lambda & & T^1_{\;1} = T^2_{\;2}=T^3_{\;3}=-p(r,t)+
\Lambda  \nonumber\\
 & T^4_{\;4} = ...=T^n_{\;n} = -p_n(r,t)+\Lambda &  
\label{set}
\eea
($\Lambda$ is the $4+n$ cosmological constant, and $p_n$ is the internal 
pressure). 

The Einstein field equations for the metric (\ref{met}) take the form
\begin{mathletters}
\beq
{\dot{A}'\over A}-{\dot{R}A'\over RA}=0
\label{fieq1}
\eeq
\beq
2{\ddot{R}\over R}+{\dot{R}^2\over R^2}+2n{\dot{R}\dot{A}\over RA}+n{\ddot{A}
\over A}+{n(n-1)\over 2}{\dot{A}^2\over A^2}-{2n\over r}{A'\over AR^2}+{n(1-n)
\over 2}{A'^2\over A^2R^2}=\Lambda -8\pi p
\label{fieq2}
\eeq
\beq
2{\ddot{R}\over R}+{\dot{R}^2\over R^2}+2n{\dot{R}\dot{A}\over RA}+n{\ddot{A}
\over A}+{n(n-1)\over 2}{\dot{A}^2\over A^2}-n{A''\over AR^2}-{n\over r}
{A'\over AR^2}+{n(1-n)
\over 2}{A'^2\over A^2R^2}=\Lambda -8\pi p
\label{fieq3}
\eeq
\beq
3{\dot{R}^2\over R^2}+{n(n-1)\over 2}{\dot{A}^2\over A^2}+3n{\dot{R}\dot{A}
\over RA}-n{A''\over AR^2}+{n(1-n)\over 2}{A'^2\over A^2R^2}-{2n\over r}
{A'\over AR^2}=\Lambda +8\pi \rho
\label{fieq4}
\eeq
\begin{eqnarray}
(n-1){\ddot{A}\over A}+(n-1)(n-2){\dot{A}^2\over A^2}+3(n-1){\dot{R}\dot{A}
\over RA}+3{\ddot{R}\over R}+3{\dot{R}^2\over R^2}-(n-1){A''\over AR^2} & 
\nonumber \\
-(n-1)(n-2){A'^2\over A^2R^2}-{2(n-1)\over r}{A'\over AR^2}=\Lambda 
-8\pi p_n &
\label{fieq5}
\end{eqnarray}
\end{mathletters}
where a dot (a prime) denotes differentiation with respect to time (radial) 
coordinate \footnote{ The system (\ref{fieq1})-(\ref{fieq5}) reduces to the 
equations given in \cite{chat4} in the case $n=1$.}. We also assume an equation 
of state of the form $p=k\rho$.

Equation (\ref{fieq1}) can be integrated and its solution is
\beq
A(r,t)=F(r)R(t)+G(t)
\label{c1}
\eeq
where $F(r)$ and $G(t)$ are arbitrary functions. 

From (\ref{fieq2}) and (\ref{fieq3}) we can obtain
\beq
A(r,t)=f(t)br^2+g(t)
\label{c2}
\eeq
where $b$ is an arbitrary constant. Comparing (\ref{c1}) and (\ref{c2}) it 
follows that 
\beq
A(r,t)=br^2R(t)+g(t)
\label{rel}
\eeq
Now using (\ref{rel}) and the equation of state in equations (\ref{fieq2})
and (\ref{fieq4}) we obtain the following system of equations:
\begin{mathletters}
\beq
\left[{k\over 2}f+{n\over 2}(n+3)+1\right]\dot{R}^2-(1+k)\Lambda R^2+
(n+2) R\ddot{R}=0
\label{com1}
\eeq
\bea
nR\ddot{g}+n[k+(k+1)(n+1)]\dot{R}\dot{g}+\left\{{\dot{R}^2\over R}[2(n+1)+
3k(n+2)]-2\Lambda (1+k)R \right.& \nonumber\\ 
\left.\mbox{}+(n+4)\ddot{R}\right\}g-2bn[n+1+k(n+2)]=0 & 
\label{com2}
\eea
\bea
[\dot{R}^2(1+3k)+2R\ddot{R}-\Lambda (1+k)R^2]g^2+{n\over 2}(1+k)(n-1)
R^2\dot{g}^2+n(3k+2)R\dot{R}g\dot{g} & \nonumber\\ 
\mbox{}+nR^2g\ddot{g}-2nb(2+3k)Rg=0 & 
\label{com3}
\eea
\end{mathletters}
where $f=(n+2)(n+3)$. In the case $n=1$, equation (\ref{com3}) reduces to an 
identity.

The metric given in (\ref{met}) will be a solution of Einstein equations
if and only if equations (\ref{com1}), (\ref{com2}) and (\ref{com3}) are 
compatible. If this is the case, equations (\ref{fieq3}),(\ref{fieq4})
and (\ref{fieq5}) act as definitions of $\rho$, $p$ and $p_n$ respectively.

The general solution to equation (\ref{com1}) is
\beq
R(t)=a\left( \exp\left[\frac{(1+k)\sqrt{2f\Lambda}}{n+2}t\right] - c\over \exp
\left(\frac{(1+k)\sqrt{\Lambda f/2}}{n+2}t\right]\right)^{2(n+2)\over f(k+1)}
\label{erre}
\eeq
where $\Lambda >0$, and $a$ and $c$ are arbitrary constants.

Without loss of generality, we take $c=1$, which corresponds to zero volume
of the 3-space at $t=0$. Replacing the expression (\ref{erre}) in (\ref{com2})
we get
\beq
\ddot g +\sqrt{2\Lambda\over f}\;\coth(2\beta t)\dot{g}-{\Lambda\over f}\;[n+3+
(n+1)\coth^2(2\beta t)]g={2(n+1)b\over a}\;[2\sinh(2\beta t)]^{-{2\over n+3}}
\label{kil}
\eeq
where $\beta =\sqrt{2\Lambda f/4(n+2)}$.
It has not been possible up to now to solve this equation neither in
the case of arbitrary values for $k$ and $n$ nor with a given $k$ and arbitrary
$n$ \footnote{ Note however that a solution
has been found in the case $k=0$, $n=1$ by Chatterjee {\em et al} 
\cite{chat4}.}. From now on we consider the case $c=0$ which
corresponds to an inflationary solution for the 3-d space with a
singularity at infinite past. The scale factor $R(t)$ takes the form
\beq
R(t)=a\exp\left(\sqrt{2\Lambda\over f}t\right)
\eeq
After replacing this equation in (\ref{com2}), we get a differential equation
for $g(t)$ that can be integrated. Its solution is
\beq
g(t)=C_1 \exp\left(\sqrt{2\Lambda\over f}t\right)+C_2 \exp\left[
-\sqrt{2\Lambda\over f}(n+2)(k+1)t\right]+{fb\over 2a\Lambda}\exp
\left(-\sqrt{2\Lambda\over f}t\right)
\label{ge}
\eeq
where $C_1$ and $C_2$ are integration constants.

Now we have to look for the conditions under which equation (\ref{com3})
is satisfied. Upon replacement of (\ref{erre}) and (\ref{ge}) in (\ref{com3})
we get the following equations in $n$:
\begin{eqnarray}
(k^2+2k+1)(-n^4-6n^3-9n^2+4n+12) & = & 0 \\
(k+1)(n^3+4n^2+n-6) & = & 0 \\
n^3[6k(1+k)+2(1+k^3)]+2n^2(3k^3+11k^2+13k+5)+2n((2k^2+5k+3) & & \nonumber\\
-2(9+21k+16k^2+4k^3) & = & 0
\end{eqnarray}
The value $n=1$ is a root of the three polynomials, irrespective of the value
of $k$. Besides, if we restrict the values of $k$ to 0 or 1/3, we see that 
there is only one root of the three polynomials that is a natural number, and 
that is again $n=1$ \footnote{ Solutions for $k=0$ and $k=1/3$ with $n=1$ 
are given in \cite{ban2}.}. \\

We conclude then that the model described by the metric (\ref{met}), the
stress-energy tensor (\ref{set}), and the equation of state $p=k\rho$ is 
integrable in the case 
of a 5-dimensional space-time for any value of $k$ (with $c=0$).
Furthermore, the cases of dust ($k=0$) and radiation
($k=1/3$) are integrable (with $c=0$) if and only if $n=1$.

Finally, let us mention that the model studied here can be generalized by 
introducing inhomogeneity in $R(t)$, and also by adding some degree of
anisotropy in the extra dimensions. Work along this lines is currently in
progress.\\

The author would like to thank H. Vucetich for helpful discussions. This
work was supported by CONICET and UNLP.

\end{document}